# MULTI-SPEAKER DOA ESTIMATION IN BINAURAL HEARING AIDS USING DEEP LEARNING AND SPEAKER COUNT FUSION

*Farnaz Jazaeri*[1] *François Grondin*[2] *Homayoun Kamkar-Parsi*[3] *Martin Bouchard*[1]

[1] School of Electrical Engineering and Computer Science, University of Ottawa, Canada
[2] Department of Electrical and Computer Engineering, Université de Sherbrooke, Canada
[3] WS Audiology, Germany

{fjaza018, bouchm}@uottawa.ca, francois.grondin2@usherbrooke.ca, homayoun.kamkarparsi@wsa.com

## ABSTRACT

For extracting a target speaker voice, direction-of-arrival (DOA) estimation is crucial for binaural hearing aids operating in noisy, multi-speaker environments. Among the solutions developed for this task, a deep learning convolutional recurrent neural network (CRNN) model leveraging spectral phase differences and magnitude ratios between microphone signals is a popular option. In this paper, we explore adding source-count information for multi-sources DOA estimation. The use of dual-task training with joint multi-sources DOA estimation and source counting is first considered. We then consider using the source count as an auxiliary feature in a standalone DOA estimation system, where the number of active sources (0, 1, or 2+) is integrated into the CRNN architecture through early, mid, and late fusion strategies. Experiments using real binaural recordings are performed. Results show that the dual-task training does not improve DOA estimation performance, although it benefits source-count prediction. However, a ground-truth (oracle) source count used as an auxiliary feature significantly enhances standalone DOA estimation performance, with late fusion yielding up to 14% higher average F1-scores over the baseline CRNN. This highlights the potential of using source-count estimation for robust DOA estimation in binaural hearing aids.

## 1. INTRODUCTION

For hearing-aid users, accurately localizing active speakers is essential for speech intelligibility and situational awareness. Direction-of-arrival (DOA) estimation enables the device to steer beamformers, suppress noise, and enhance conversational cues. However, real-world listening environments such as restaurants and meeting rooms pose significant challenges: multiple simultaneous talkers, reverberation, and background noise that can degrade localization performance.

Classical direction-of-arrival (DOA) estimation for acoustic sources typically relies on time- and phase-difference cues. Methods such as GCC-PHAT, MUSIC/ESPRIT, and SRP-PHAT remain effective in simple conditions, but degrade under reverberation and overlapping sources [1, 2, 3, 4, 5, 6]. While subspace and beamforming approaches extend to multi-source cases, they struggle with realistic noisy environments and head-related transfer function (HRTF) coloration.

Deep learning (DL) methods have been able to provide improved robustness by learning spatial–spectral features directly from data. Such systems have shown strong performance for single-source localization [7, 8, 9]. For multi-sources DOA estimation, deep learning CRNN models leveraging convolutional neural networks and recurrent neural networks have shown to be competitive [7, 10, 11, 12, 13]. In most cases, these systems treat DOA estimation as a multi-class multi-label classification problem [14]. Spectral phase differences and magnitude ratios between microphone signals have often been proposed as simple features to provide spatial and spectral cues [7, 10, 11, 15].

For deep learning DOA estimation systems, including source-count information could potentially be beneficial to improve the performance [15, 16, 17]. In [15], a deep learning model was used for the dual-task of source counting and single-source DOA estimation. It was shown that the dual task helped to improve the source counting performance. However, the improvement on the DOA estimation was not directly reported, i.e., the assessment of the DOA estimation was done comparing results from the dual-task model to a classic method.

Whether or not including source count information can improve multi-sources DOA estimation (or even single-source DOA estimation) is therefore an open question. This paper aims to provide some answers to this question. First, the use of dual-task training for joint multi-sources DOA estimation and source counting is evaluated. We then evaluate using the source count as an auxiliary feature in a DOA estimation system, where the number of active sources (0, 1, or 2+) is integrated into a CRNN architecture through early, mid, and late fusion strategies.

Our contributions are: (i) investigation of two architec-tures for improving multi-sources DOA estimation (dual task and source count as input feature); (ii) systematic evaluation of early/mid/late fusion for integrating source count information into DOA networks; and (iii) assessment of how the resulting models obtained from synthetic data (based on head-related impulse responses (HRIRs)) generalize to real recordings from restaurants and coffee shops or controlled laboratory recordings.

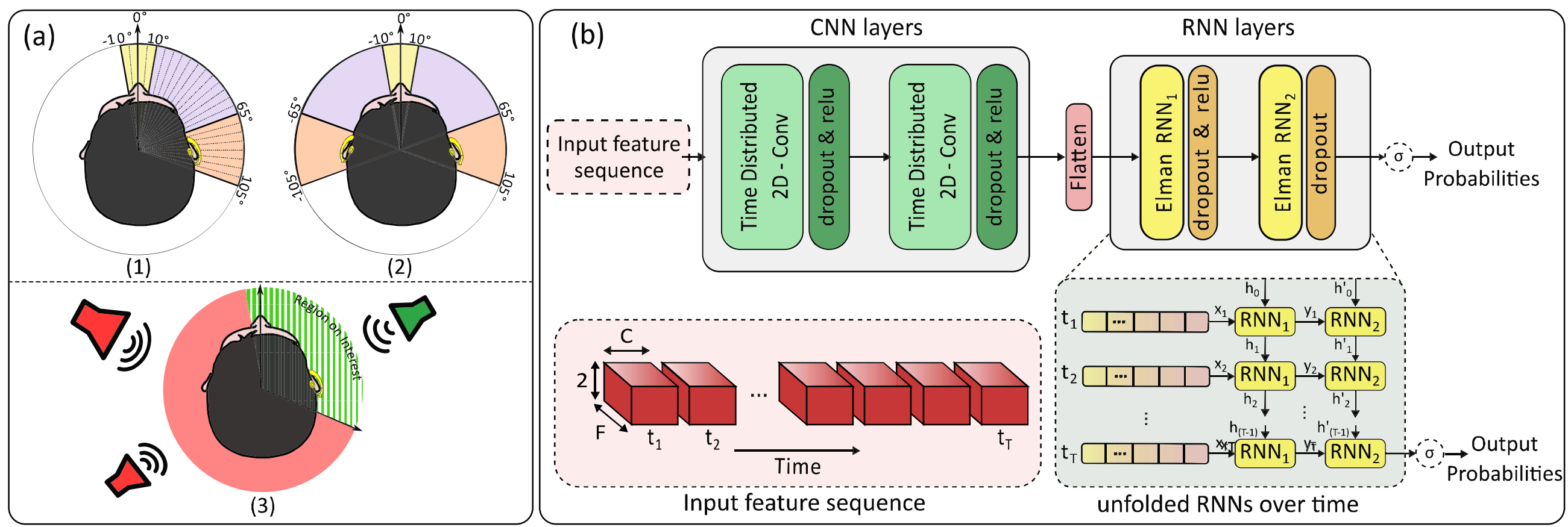

**Fig. 1**: (a) DOA estimation in different regions + ROI concept. (b) baseline CRNN architecture.

## 2. METHODOLOGY

### 2.1. Problem Formulation

We estimate direction-of-arrival (DOA) from a binaural 2-microphone behind-the-ear (BTE) binaural hearing-aid, using three microphone signals: front/rear microphones on the local device, and front microphone from the opposite-side device. Taking the view of the right-side device in Fig.1(a), multi-sources DOA estimation with a $5^\circ$ angular resolution is cast as a multi-class multi-label classification over a region of interest (ROI) with either 16 classes for $[-10^\circ, 65^\circ]$ or 24 classes for $[-10^\circ, 105^\circ]$. For each class $d \in \{1, \ldots, D\}$, where $D \in \{16, 24\}$, the network outputs a logit $z_d$, with posterior probability,

$$p_d = \sigma(z_d) \in [0, 1], \tag{1}$$

allowing multiple simultaneous active DOAs.

### 2.2. Feature Extraction

Features are derived from subband analysis of the three microphone signals using 200 ms frames with 50% overlap, for 11 bands up to 5 kHz. We compute inter-channel phase differences (IPD) from cross-power spectral density, and inter-channel level ratios (ILR) normalized to a reference microphone. These components are concatenated into a feature map:

$$\mathbf{F} \in \mathbb{R}^{T \times F \times C \times 2}, \tag{2}$$

where $T \in \mathbb{N}$ is the number of frames, $F \in \mathbb{N}$ is the number of subbands, and $C \in \mathbb{N}$ is the number of intra-microphone phase differences or magnitude ratios. The sequence $\mathbf{F}$ is then fed to the CRNN architecture.

### 2.3. Baseline DOA Model

Each frame slice is encoded by two convolutional layers (CNNs) that extract spatial–spectral patterns from phase differences and level ratios input features. The sequence of CNN outputs $\{\mathbf{h}_{\text{CNN}}(t)\}$ is modeled by a recurrent layer (RNN) to capture temporal context, which is critical under noise and reverberation.

The RNN outputs are passed to independent sigmoids, yielding class probabilities $p_d(t) \in [0, 1]$ for each DOA $d$. Multi-label detection allows to detect zero, one or several active sources per frame. The network processes input sequences of length $T = 10$ (1 s). Only the last frame $t = T$ contributes to the loss, and we denote $y_d(T)$ and $p_d(T)$ as $y_d$ and $p_d$, respectively, for clarity:

$$L_{\text{DOA}} = -\sum_{d \in \mathcal{D}} \left[ y_d \log p_d + (1 - y_d) \log (1 - p_d) \right]. \tag{3}$$

A compact CRNN is used to balance performance and efficiency, with $\approx 38$k parameters. The architecture of the baseline CRNN is shown in Fig. 1(b).

### 2.4. Dual-Task Learning

Results in [15] showed that joint modeling of DOA and source count improved performance of source count and led to single-source DOA estimation which outperformed a classic method. Inspired by this, for multi-sources DOA estimation we extend the baseline system with a Concurrent Speaker Detection (CSD) branch, as shown in Fig. 2(a). The CSD head classifies the source count into three categories:

$$y_{\text{CSD}} \in \{0, 1, 2+\}. \tag{4}$$

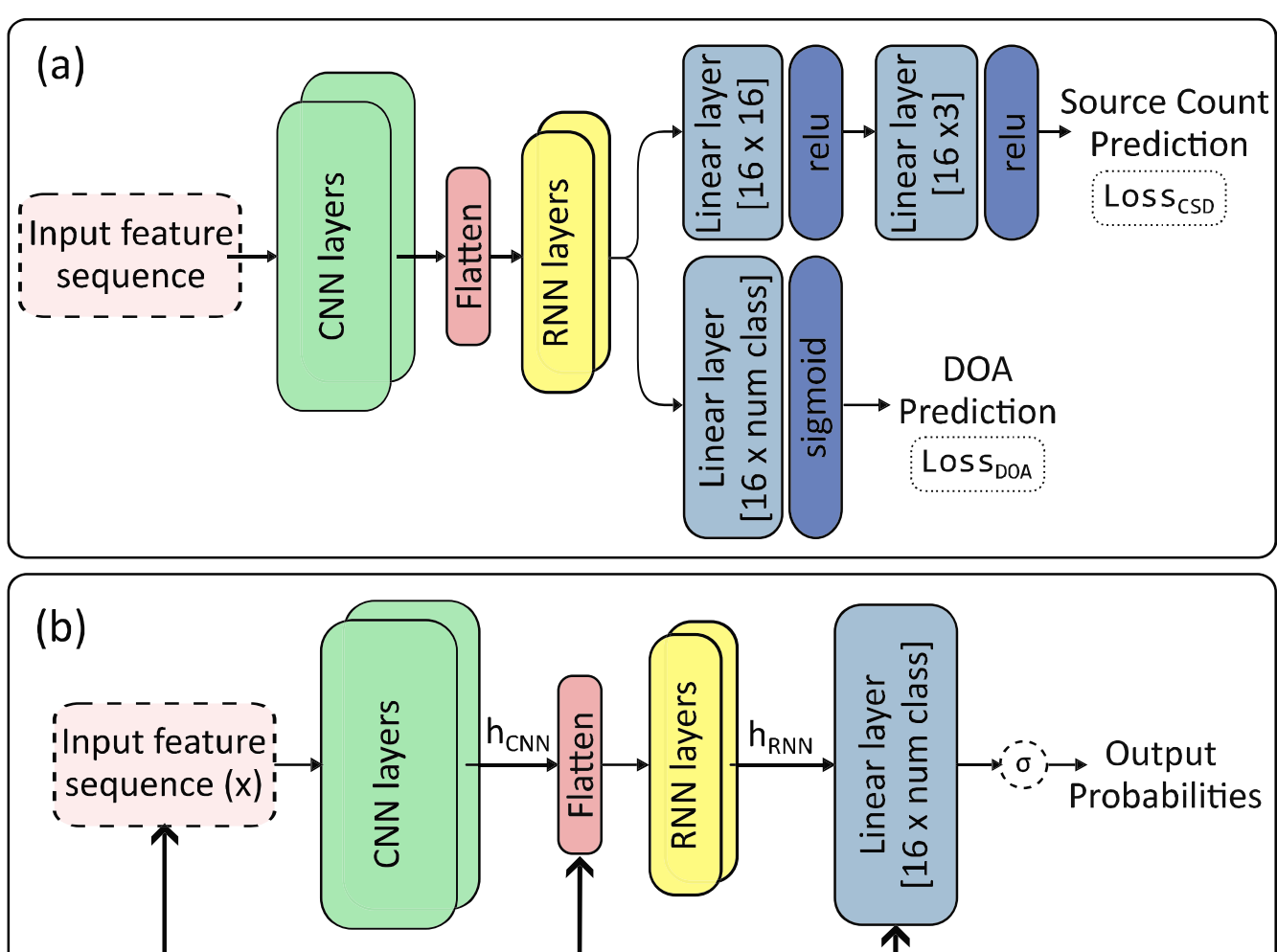

**Fig. 2**: (a) Dual-task CRNN for joint DOA and source-count prediction. (b) Fusion of source-count features at input, intermediate, or output stage to guide DOA estimation.

Note that by default we assume that the source count is performed in the same ROI as the DOA estimation, e.g., the shaded green region in Fig. 1(a).

The CSD branch shares the CRNN encoder with the DOA task and applies a small classifier, using cross-entropy:

$$L_{\text{CSD}} = -\sum_{c\in\{0,1,2+\}} q_c \log \hat{q}_c, \tag{5}$$

where $q_c \in \{0,1\}$ is the one-hot label and $\hat{q}_c \in [0,1]$ the predicted probability. As with DOA, only the final frame in each sequence contributes to the loss. The two tasks are trained jointly with a weighted objective:

$$L = \alpha L_{\text{DOA}} + (1-\alpha) L_{\text{CSD}}, \quad \alpha \in [0,1], \tag{6}$$

which balances localization accuracy and speaker counting.

### 2.5. Fusion of Speaker Count into DOA

The use of explicit speaker-count information to directly support DOA estimation was next evaluated, as shown in Fig. 2(b). Knowing the number of active sources within the ROI constrains the task: if a single source is present, the network should assign high probability to only one DOA, whereas in multi-speaker cases it must distribute probability mass across several classes. Alternatively, another approach to use the speaker-count information could be switch between different models or model sub-nets, e.g., modules for single-sources scenarios or multi-sources scenarios, but this was not investigated in this work.

We tested three fusion strategies that integrate the count embedding $s$ (one-hot over $\{0,1,2+\}$) at different stages of the network, to probe whether count cues help most at feature learning, temporal modeling, or final decision: **Early fusion** augments the raw input $\mathbf{x}$ with $s$, so that the CNN encoder can learn feature representations conditioned on the number of active sources. **Mid fusion** inserts $s$ after the CNN block, biasing the intermediate spatial–spectral features $h_{\text{CNN}}$ toward representations consistent with the source count. **Late fusion** appends $s$ after the RNN block, guiding the final temporal representation $h_{\text{RNN}}$ to enforce source-count constraints before classification. Formally, the fused representation is:

$$\mathbf{h}_{\text{fused}} = \begin{cases} \text{Concat}(\mathbf{x}, s), & \text{Early fusion}, \\ \text{Concat}(\mathbf{h}_{\text{CNN}}, s), & \text{Mid fusion}, \\ \text{Concat}(\mathbf{h}_{\text{RNN}}, s), & \text{Late fusion}. \end{cases} \tag{7}$$

To assess the potential of this approach, we conducted oracle experiments with ground-truth counts. These quantify the upper bound achievable if a reliable CSD module were available. In all cases, the count corresponds to the number of active sources in the ROI (i.e., not over 360°), consistent with the DOA labels.

## 3. EXPERIMENTS

### 3.1. Datasets

Training relied on synthetic mixtures generated by convolving TIMIT speech with head-related impulse responses (HRIRs) from multiple WS Audiology behind-the-ear (BTE) 2-microphone binaural hearing aid devices, measured in both anechoic and reverberant rooms (with RT60 reverberation time from 0.1 to 0.6 sec). Mixtures were produced for 0, 1, or 2 active sources over 360°. Multichannel diffuse noise was also added at SNRs of 5, 10, and 15 dB to the training dataset. This procedure yields ∼26 h of mixtures (5,550 clips of approx. 17 sec. each).

Our evaluation used real binaural recordings: (i) restaurants and coffee shops recordings of conversations with 2–4 speakers with natural noise conditions, and (ii) laboratory small room recordings with 1–3 speaker(s) at known DOAs, with controlled noise levels. Table 1 provides an overview.

### 3.2. Implementation Details

Models were implemented in PyTorch Lightning and trained on an NVIDIA H100 GPU. Training used the Adam optimizer with an initial learning rate of $10^{-2}$ , decayed at predefined milestones. A batch size of 512 was employed, and each model converged within 20 minutes for 25 epochs. Regularization was applied through a dropout rate of 0.1, and early stopping was triggered based on validation loss. Evaluation metrics included region-wise F1-scores (frontal, diagonal, lateral, as shown in Fig. 1(a)) computed by pooling detections across left and right hearing aids. No additional data augmentation beyond diffuse noise mixing was applied.

**Table 1**: Overview of datasets used for training, validation, and testing.

| Dataset | Type | Source / Content | Environment | Duration/clips | Usage |
|---|---|---|---|---|---|
| Synthetic mix (TIMIT+HRIR) | Synthetic | TIMIT speech convolved with HRIRs + diffuse noise | Anechoic & reverberant rooms | ∼26 h (5,550) | Train/Val |
| Rest./coffee shops record. | Real | Conversations (2–4 speakers + background) | Public restaurants | ∼46 m (24) | Test |
| Lab. recordings | Real | 1–3 speakers with labels | Quiet controlled room | ∼31 m (13) | Test |

**Table 2**: DOA estimation results on Restaurants / coffee shops recordings and Lab. recordings datasets (24-class setup).

| Method | Restaurants / coffee shops recordings | | | | Lab. recordings | | | |
|---|---|---|---|---|---|---|---|---|
| | Frontal F1 | Diagonal F1 | Lateral F1 | Avg | Frontal F1 | Diagonal F1 | Lateral F1 | Avg |
| Baseline – CRNN | 0.84 | 0.73 | 0.32 | 0.63 | 0.70 | 0.79 | 0.77 | 0.75 |
| Dual-Task training | 0.83 | 0.72 | 0.29 | 0.61 | 0.76 | 0.75 | 0.76 | 0.76 |
| Oracle count – Early | 0.88 | 0.74 | 0.40 | 0.67 | 0.80 | **0.82** | 0.76 | 0.79 |
| Oracle count – Mid | **0.89** | 0.75 | 0.52 | **0.72** | **0.83** | 0.80 | 0.81 | 0.81 |
| Oracle count – Late | 0.86 | **0.76** | **0.51** | 0.71 | 0.80 | **0.82** | **0.88** | **0.83** |

**Table 3**: DOA estimation results on Restaurants / coffee shops recordings and Lab. recordings datasets (16-class setup).

| Method | Restaurants / coffee shops recordings | | | Lab. recordings | | |
|---|---|---|---|---|---|---|
| | Frontal F1 | Diagonal F1 | Avg | Frontal F1 | Diagonal F1 | Avg |
| Baseline – CRNN | 0.87 | 0.75 | 0.81 | 0.72 | 0.72 | 0.72 |
| Dual-Task training | 0.88 | 0.75 | 0.82 | 0.76 | 0.75 | 0.76 |
| Oracle count – Early | **0.95** | 0.77 | 0.86 | 0.85 | 0.83 | 0.84 |
| Oracle count – Mid | 0.93 | 0.76 | 0.85 | 0.83 | 0.78 | 0.81 |
| Oracle count – Late | 0.94 | **0.84** | **0.89** | **0.86** | **0.86** | **0.86** |

### 3.3. Results

For the dual-task architecture, the value of $\alpha$ in (7) was adjusted between 0.95 and 0.99 for each experiment to produce the highest F1-score. The performance of source counting at testing was greatly improved with the dual-task, with F1-scores increasing from values below 0.2 to above 0.6 for the 2+ class. But the results in Tables 2 and 3 show that the dual-task architecture did not provide improvement for DOA estimation, compared to the baseline. This is in line with the results in [15], where the only improvement observed for DOA estimation from dual-task architecture was when comparing to a classic method. A possible reason for the inability to improve multi-source DOA estimation with the dual-task architecture is that multi-label DOA outputs already encode implicit count information, making the CSD branch redundant.

However, Tables 2 and 3 show that including ROI specific oracle source count as auxiliary information yielded consistent F1-score gains across frontal, diagonal, and lateral regions, with the largest improvements observed in diagonal and lateral regions under challenging multi-speaker conditions. In particular, *late fusion* provided the strongest benefits, improving average F1-scores by approximately 8–9% in the 24-class setup and 8–14% in the 16-class setup compared to the baseline CRNN model.

Overall, despite being trained entirely on synthetic HRIR-based mixtures, the proposed models generalized well to real restaurants/coffee shops and laboratory recordings, except for the lateral region in restaurants/coffee shops recordings.

## 4. CONCLUSION

This work evaluated whether source-count information can improve DOA estimation in binaural hearing aids. Dual-task training with DOA estimation and source-count did not improve DOA performance, suggesting that multi-label DOA outputs already encode implicit count information. However, explicit oracle source-count as auxiliary information substantially improved multi-sources DOA estimation. Performance with estimated counts remains to be evaluated. Results were based on a CRNN model with intra-mic. phase and magnitude features, and may not generalize to other systems. The findings highlight that the development of practical ROI-specific speaker counting modules holds strong potential for enhancing DOA estimation performance in hearing aids devices.